\documentclass[conference]{IEEEtran}
\IEEEoverridecommandlockouts
\usepackage{amsmath,amssymb,amsfonts}
\usepackage{accents}
\usepackage{algorithm}
\usepackage{algpseudocode}
\usepackage{bbm}
\usepackage{hyperref}
\usepackage{fancyhdr}
\usepackage{graphicx}
\usepackage{setspace}
\usepackage{textcomp}
\usepackage{xcolor}
\usepackage{multicol}
\usepackage{placeins}
\usepackage{booktabs}
\usepackage{subcaption}
\usepackage{float}
\usepackage[backend=biber,style=ieee,sorting=none,firstinits=true]{biblatex}
\newcommand{\XB}{\boldsymbol{X}}
\newcommand{\xB}{\boldsymbol{x}}
\newcommand{\yB}{\boldsymbol{y}}
\newcommand{\betaB}{\boldsymbol{\beta}}
\newcommand{\varepsilonB}{\boldsymbol{\varepsilon}}
\newcommand{\stildeB}{\tilde{\mathbf{s}}}
\newcommand{\PhiB}{\boldsymbol{\Phi}}

\newcommand{\XK}{\boldsymbol{\protect\accentset{\circ}{X}}}
\newcommand{\xK}{\boldsymbol{\protect\accentset{\circ}{x}}}
\newcommand{\xk}{\protect\accentset{\circ}{x}}
\newcommand{\Xwtb}{\widetilde{\boldsymbol{X}}}
\newcommand{\xwtb}{\widetilde{\boldsymbol{x}}}
\newcommand{\CN}{\mathcal{C}\mathcal{N}}
\newcommand{\Ibb}{\mathbbm{1}}

\newcommand{\degreeSign}{\circ}
\newcommand{\iid}{\text{i.i.d.}}
\newcommand{\SNR}{\mathrm{SNR}}
\newcommand{\imagunit}{\mathrm{i}}
\newcommand{\hop}{\ensuremath{^{\text{H}}}}

\begin{document}
% -------------------------------------
\setstretch{0.94} % Consider removing this if it conflicts with IEEE formatting
\title{\Huge
    FDR Control for Complex-Valued Data with Application in Single Snapshot Multi-Source Detection and DOA Estimation
    %FDR Control in High-Dimensional Complex Space: Single Snapshot DOA Estimation Using the Complex Terminating Random Experiment Selector
    %FDR Control in High-Dimensional Complex Space\\
    %The Complex Terminating Random Experiment Selector
    %
    \thanks{The first author is supported by the federal German BMBF Clusters4Future
    initiative curATime, within the curAIsig project under grant number 03ZU1202MA.
    The second author is supported by the LOEWE initiative (Hesse, Germany) within the emergenCITY center.
    The third author is supported by the ERC Starting Grant ScReeningData under grant number 101042407.
    \textcopyright~2025 IEEE. Personal use permitted.
    Final version: ieeexplore.ieee.org/document/10889705
    }
}

\author{
    \IEEEauthorblockN{Fabian Scheidt, Jasin Machkour, Michael Muma}
    \IEEEauthorblockA{
        \emph{Robust Data Science Group}\\
        \emph{Technische Universität Darmstadt}\\
        64283, Darmstadt, Germany \\
        \{fabian.scheidt, jasin.machkour, michael.muma\}@tu-darmstadt.de
    }
}

\maketitle

% ------------------------------------------------------------------------------------
\begin{abstract}
% Your abstract content goes here.
% -----------------------------------
False discovery rate (FDR) control is a popular approach for maintaining the integrity of statistical analyses, especially in high-dimensional data settings, where multiple comparisons increase the risk of false positives. FDR control has been extensively researched for real-valued data. However, the complex data case, which is relevant for many signal processing applications, remains widely unexplored. We therefore present a fast and FDR-controlling variable selector for complex-valued high-dimensional data. The proposed Complex-Valued Terminating-Random Experiments (\emph{CT-Rex}) selector controls a user-defined target FDR while maximizing the number of selected variables. This is achieved by optimally fusing the solutions of multiple early terminated complex-valued random experiments. We benchmark the performance in sparse complex regression simulation studies and showcase an example of FDR-controlled compressed-sensing-based single snapshot multi-source detection and direction of arrival (DOA) estimation. The proposed work applies to a wide range of research areas, such as DOA estimation, communications, mechanical engineering, and magnetic resonance imaging, bridging a critical gap in signal processing for complex-valued data.
\end{abstract}

% ------------------------------------------------------------------------------------

% ------------------------------------------------------------------------------------
\section{Introduction}
% Your introduction content goes here.
% -----------------------------------
False discovery rate (FDR) control is widely used to maximize the number of true positive detections while constraining the expected proportion of false positives. It is particularly beneficial in high-dimensional data scenarios, where the risk of spurious detections increases due to the large number of hypotheses that are tested. Ensuring FDR control is crucial for preserving the reliability and interpretability of statistical analyses in such settings. While substantial research has focused on developing FDR-controlling methods for real-valued data (for low- and high-dimensions \cite{benjamini1995control, benjamini2001control, candes2018panning, machkour2021terminating}), the extension to complex-valued data, a key requirement in many signal processing applications, remains relatively underexplored. This work contributes to the existing literature on complex-valued statistics, which has focused on probability theory fundamentals, estimation methodologies, and challenges that relate to non-circularity \cite{adali2011complex, zoubir2018robust, schreier2010statistical, ollila2011complex, picinbono1994circularity, eriksson2010essential, van1995multivariate}. To the best of our knowledge, we present the first FDR-controlling method for high-dimensional complex-valued data. We build upon the recently proposed Terminating-Random Experiments (\emph{T-Rex}) Selector \cite{machkour2021terminating, machkour2022false, machkour2023false, TLARSpackage, TRexSelectorpackage, machkour2024high, scheidt2023solving} by introducing its complex-valued counterpart. The \emph{CT-Rex} optimally fuses the solutions of multiple early-terminated complex-valued random experiments, where computer-generated dummy variables compete with real data in a forward variable selector. As an important building block of our method, we implemented the Complex Terminated-Least-Angle-Regression (\emph{CT-LARS}) algorithm, which serves as a forward selector. Our research aims to establish robust tools for complex-domain FDR control in applications such as compressed-sensing-based DOA estimation \cite{malioutov2005sparse, tabassum2016single, tabassum2018sequential}, mechanical engineering \cite{brunton2016discovering, graff2020reduced, jovanovic2014sparsity}, and magnetic resonance imaging \cite{lustig2007sparse}, bridging a critical gap in signal processing and engineering literature.\\
\emph{Organization}: Sec.~\ref{sec:CTRex_selector} presents the \emph{CT-Rex} Selector. Sec.~\ref{sec:TRex_Theory_Revisited} details the most important changes resulting from the complex-data case. 
Sec.~\ref{sec:Simulations} presents simulation results for sparse complex regression and DOA estimation.
Sec.~\ref{sec:Conclusions} concludes the paper.

% ------------------------------------------------------------------------------------
\section{THE COMPLEX \emph{T-REX} SELECTOR}
\label{sec:CTRex_selector}
\begin{figure}
    		\centering
            \includegraphics[width=\linewidth]{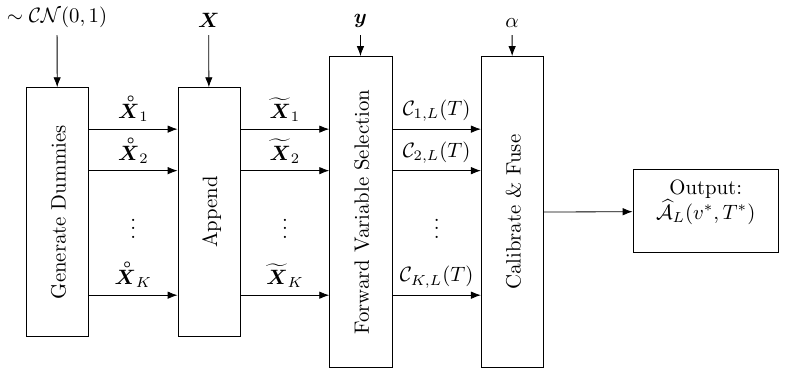}
    	    \caption{Sketch of the \emph{CT-Rex} selector.
            }
         \label{fig:CTRex_scheme}
\end{figure}

% Your section content goes here.
% -----------------------------------
The complex \emph{T-Rex} (\emph{CT-Rex}) selector extends the real-valued \emph{T-Rex} selector framework \cite{machkour2021terminating, machkour2022false, machkour2023false, TLARSpackage, TRexSelectorpackage, machkour2024high, scheidt2023solving} to the complex data domain.
It controls a user-defined target FDR while maximizing the number of selected variables by modeling and fusion of multiple early terminated complex-valued random experiments, in which pseudo-random number generated dummy variables compete with the original variables.
The \emph{CT-Rex} does not require any user-defined tuning. Its inputs are:
\begin{enumerate}
    \item{The predictor matrix $\XB = [\xB_{1} \dots \xB_{p}]$, whose $p$ predictors each contain $n$ samples, i.e., $\XB \in \mathbb{C}^{n \times p}$.}
    \item{The response vector $\yB = [y_{1} \dots y_{n}]^\top$, $\yB \in \mathbb{C}^{n}$.}
    \item{The user-defined target FDR level $\alpha \in [0, 1]$.}
\end{enumerate}
A sketch of the \emph{CT-Rex} selector algorithm is provided in Fig.~\ref{fig:CTRex_scheme}.
The main contribution of this work is to extend the \emph{T-Rex} framework, so that all processing steps such as the dummy generation and forward variable selection are performed in the complex number domain.
The \emph{CT-Rex} steps are briefly summarized in the following:\\
\emph{Step~1 (Generate Dummies)}: A set of $K > 1$ dummy matrices $\{\XK_{k} = [\xK_{k,1} \dots \xK_{k,L}]\}_{k=1}^{K}$, each with $L$ dummy variables, is generated from a circularly symmetric white probability distribution with the condition of finite mean and nonzero finite variance, e.g., the complex standard normal $\CN(0, \sigma^2\mathbf{I})$.\\
\emph{Step~2 (Append Complex Dummies)}: The original predictor matrix $\XB$ is augmented $K$ times and a set of enlarged predictor matrices $ \{ \Xwtb_{k} = [\boldsymbol{X} \,\, \XK_{k} ] \}_{k=1}^{K} $ is formed.
This establishes an auxiliary tool for FDR-controlled variable selection in which original and dummy variables compete for selection.\\
\emph{Step~3 (Complex Valued Forward Variable Selection)}: A complex-valued forward variable selector is applied to a set of tuples $\{ (\Xwtb_{k}, \yB) \}_{k=1}^{K} $ and terminates each of these $K$ random experiments when $T = 1$ dummy variable enters the respective active sets.
Let
\begin{equation}
    \yB = \XB \betaB + \varepsilonB
    \label{eq:complex_lin_regression}
\end{equation}
be a complex-valued linear model with a complex sparse support vector $\betaB \in \mathbb{C}^{p}$, with cardinality $\| \betaB \|_{0} = s < p$, and an error term $\varepsilonB \sim \CN(0, \sigma^{2} \mathbf{I})$.
For \emph{CT-Rex} we developed a terminated-LARS algorithm for the complex domain (see Algorithm 1) building upon \cite{efron2004least, graff2020reduced, TRexSelectorpackage, TLARSpackage}.\\
\emph{Step~4 (Calibrate \& Fuse)}: The outcome of each complex-valued random experiment is a set of candidate variables $\{ \mathcal{C}_{k,l}(T) \}_{k=1}^{K}$ from which the $T$ dummies are removed, and the relative occurrence of each variable is computed as:
\begin{equation}
    \Phi_{T, L}(j) = \begin{cases}
        \frac{1}{K} \sum_{k=1}^{K} \Ibb_{k}(j, T, L), &T \geq 1 \\
        0, &T = 0
    \end{cases} \, .
\end{equation}
The indicator function $\Ibb_{k}(j, T, L)$ equals one if the $j$th variable was selected in the $k$th complex-valued random experiment.
Then, a conservative estimate of the false discovery proportion (FDP), is computed and compared against the user-defined FDR threshold $\alpha$.
If the threshold is not exceeded, the dummy count is incremented, i.e., $T \gets T + 1$, and the next iteration of complex forward variable selection is computed until the FDP exceeds the target FDR $\alpha$.
Following this procedure, the \emph{CT-Rex} automatically determines the parameters $T, L$, and $v$, s.t., the FDR is controlled at the user-defined target level (see Theorem 1 in \cite{machkour2021terminating}).\\
\emph{Step~5 (Output)}: The active set, i.e.,
\begin{equation}
    \widehat{\mathcal{A}}_{L} = \{ j: \Phi_{T^{*}, L}(j) > v^{*} \}
\end{equation}
with $v^{*}$ and $T^{*}$, denoting the optimal values for the voting level $v$ and the included dummies $T$, which maximize the number of selected variables (see Theorem 3 in \cite{machkour2021terminating}).

% ------------------------------------------------------------------------------------
\section{MAIN CHANGES FOR THE COMPLEX DOMAIN}
\label{sec:TRex_Theory_Revisited}
% Your section content goes here.
% -----------------------------------
Two main elements of the \emph{CT-Rex} are modified compared to the real-valued \emph{T-Rex} selector. These are the forward variable selection and the dummy generation.
\subsection{Complex Terminating Least Angle Regression (CT-LARS)}
\label{sec:CTLARS}
Within each $k$th random experiment, the \emph{CT-Rex} requires an early terminating version of a complex least-angle regression (\emph{CT-LARS}) selector.
Each terminating learner then determines a candidate set $\{ \mathcal{C}_{k,l}(T) \}_{k=1}^{K}$ according to Algorithm \ref{alg:ct_lars}, which is further processed according to step \emph{Step~4 (Calibrate \& Fuse)} of the \emph{CT-Rex}.
\begin{algorithm}
\caption{Complex Terminating LARS (\emph{CT-LARS})}
\label{alg:ct_lars}
\begin{algorithmic}
        \State \textbf{Input:} State information of $k$th random experiment.
        \State \quad Complex input matrix: $\widetilde{\XB}_{k} = [\XB \,\, \XK_{k}] \in \mathbb{C}^{n \times (p + L)}$.
        \State \quad Complex response vector: $\yB \in \mathbb{C}^n$.
        \State \quad Maximum number of dummy variables allowed: $T$.
        \State \textbf{Ensure:} Set of indices for dummy variables $\mathcal{D}_{k}$ is known.
            \State \quad Standardized $\widetilde{\XB}_{k}$ (zero-mean, unit-variance columns).
            \State \quad Centered $\yB$ (zero-mean).
        \If{$T = 1$}
            \State Set:
                iteration index $\tau = 0$\\
                \quad \quad $\widehat{\betaB}_{k,0} = \mathbf{0}$, $\mathcal{A}_{k,0} = \varnothing$, $\mathbf{r}_{k,0} = \yB$, $t = 0$.
        \Else
            \State Continue with:
                $\widehat{\betaB}_{k-1,\tau}, \mathcal{A}_{k-1,\tau}, \mathbf{r}_{k-1,\tau}, t$.
        \EndIf
        \State \textbf{Processing:}
        \While{$\|\mathcal{A}_{k,\tau}\|_{0} < \min(n, p+L)$ \textbf{and} $t < T$}
            \State Increment iteration index: $\tau \gets \tau+1$.
            \State Compute complex correlations:
                $\mathbf{c}_{k,\tau} = \widetilde{\XB}_{k}\hop \mathbf{r}_{k,\tau-1}$.
            \State Find most correlated predictor:\\
                \quad \quad $j^{*} = \arg\max_{j} |\mathbf{c}_{k,\tau}|$.
            \State Add index $j^{*}$ to active set:\\
                \quad \quad $\mathcal{A}_{k,\tau} = \mathcal{A}_{k,\tau-1} \cup j^{*}$.
            \If{$j^{*} \in \mathcal{D}_{k}$} $t \gets t + 1$.
            \EndIf
            \State Form signum-aligned active variables:\\
                \quad \quad $s_{k,\tau,j} = \mathrm{sgn}_{\mathbb{C}}(c_{k,\tau,j})$, $\quad j \in \mathcal{A}_{k,\tau}$,\\
                \quad \quad $\widetilde{\XB}_{\mathcal{A}_{k,\tau}} = [s_{k,\tau,j} \, \mathbf{\xwtb}_{k,j}: j \in \mathcal{A}_{k,\tau}]$,
                where $\mathbf{\xwtb}_{k} \in \Xwtb_{k}$.
            \State Create equi-angular direction of travel:
            \begin{align*}
                L_{\mathcal{A}_{k,\tau}} &= \left(\mathbf{1}^{\top} (\widetilde{\XB}_{\mathcal{A}_{k,\tau}}\hop \widetilde{\XB}_{\mathcal{A}_{k,\tau}})^{-1} \mathbf{1}\right)^{-1/2}, \\
                \mathbf{w}_{k,\tau} &= (\widetilde{\XB}_{\mathcal{A}_{k,\tau}}\hop \widetilde{\XB}_{\mathcal{A}_{k,\tau}})^{-1} L_{\mathcal{A}_{k,\tau}} \mathbf{1}, \\
                \mathbf{u}_{k,\tau} &= \widetilde{\XB}_{\mathcal{A}_{k,\tau}} \mathbf{w}_{k,\tau}.
            \end{align*}
            \State Compute correlation with direction of travel: \\ \quad \quad $\mathbf{g}_{k,\tau} = \widetilde{\XB}_{k}\hop \mathbf{u}_{k,\tau}$.
            \State Compute step size $\gamma_{k,\tau}$: according to step (2.6) in \cite{graff2020reduced}.
            \State Update coefficients:
                $\widehat{\betaB}_{\mathcal{A}_{k,\tau}} = \widehat{\betaB}_{\mathcal{A}_{k,\tau-1}} + \gamma_{k,\tau} \mathbf{w}_{k,\tau}$.
            \State Update residual:
                $\mathbf{r}_{k,\tau} = \mathbf{r}_{k,\tau-1} - \gamma_{k,\tau} \mathbf{u}_{k,\tau}$.
        \EndWhile
        \State \textbf{Output:} $\widehat{\betaB}_{k}$ in $\mathcal{C}_{k,l}(T)$.
\end{algorithmic}
\end{algorithm}

\subsection{Complex Valued Dummy Generation}
\label{sec:CmplxDummyGen}
The \emph{T-Rex} dummy generation (see Theorem 2 of \cite{machkour2021terminating}) extends straightforwardly to the complex domain by generating \iid \ dummies for each random experiment, from a circularly symmetric standard complex Gaussian distribution, i.e.,
\begin{equation}
    \xK_{l} = [\xk_{1, l} \dots \xk_{n, l}], \mathrm{with \,} \xk_{i, l} \sim \CN(0, 1) 
\end{equation}
for $i = 1, \dots, n$, $l = 1, \dots, L$.
%Dummies and Null variables are not related to the response and the higher the dummy number the higher the probability of including a dummy rather than a null in the next selection step.
%Consequently, only the number of dummies within the enlarged predictor matrices is relevant for the forward selection process.
For $n \rightarrow \infty$, and following the same steps as in \cite{machkour2021terminating}, we can show that the conditions for the Lindeberg-Feller theorem in complex space hold.
As a result, the dummies can be drawn from any univariate complex probability distribution with finite expectation and finite non-zero variance.
Note that, while the dummy generation theorem is asymptotic (i.e., $n \rightarrow \infty$), the FDR control property of the \emph{CT-Rex}, which is independent of the dummy generation theorem, is a finite sample result.

% ------------------------------------------------------------------------------------
\section{SIMULATION STUDIES}
\label{sec:Simulations}
% Your section content goes here.
% -----------------------------------
This section presents numerical simulation results which give insights into the performance of the \emph{CT-Rex} selector.
In Sec.~\ref{sec:complexRegression} we analyze a sparse linear complex regression scenario.
In Sec.~\ref{sec:single_snap_DOA} we examine a compressed sensing single snapshot direction of arrival (DOA) estimation problem.

% -------------------------------------------------------------------------------
\subsection{Complex-Valued Sparse Linear Regression}
\label{sec:complexRegression}
\begin{figure}[ht]
    \centering
    \subfloat[FDR-control property (Target FDR $\alpha = 10\%$).\label{fig:complex_regression_fdr_plot}]{
        \includegraphics[width=0.8\linewidth]{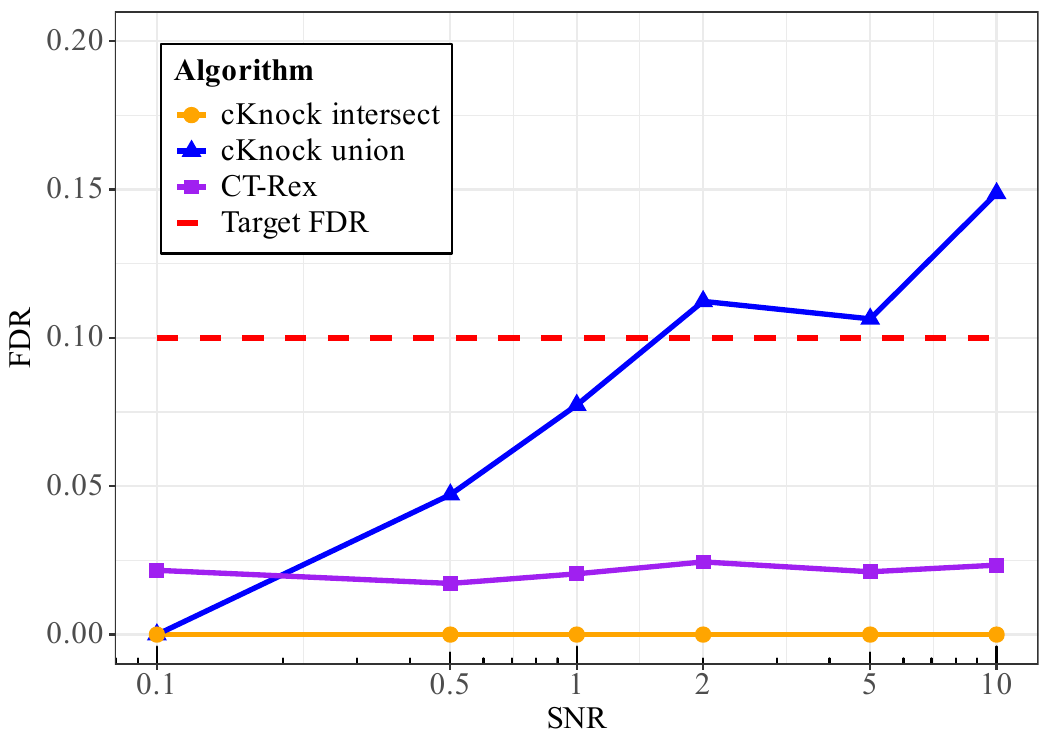}
    }
    \\
    \centering
    \subfloat[TPR performance.\label{fig:complex_regression_tpr_plot}]{
        \includegraphics[width=0.8\linewidth]{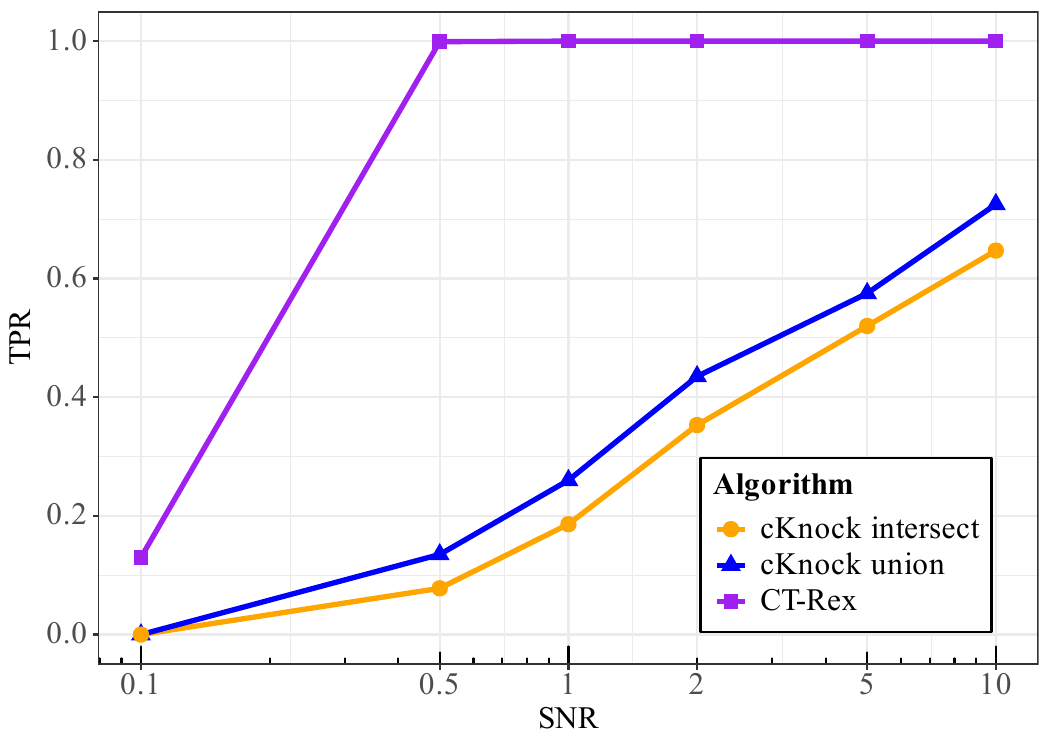}
    }
    \caption{Complex Linear Regression for varying SNR.}
    \label{fig:complex_regression_combined}
\end{figure}

In the following, we consider complex sparse linear regression, i.e., the data follows Eq.~(\ref{eq:complex_lin_regression}).
We compare the performance of the \emph{CT-Rex} selector with a complex-valued wrapped \emph{Model-X+} Knockoff selector (\emph{cKnock}) inspired by \cite{candes2018panning}.
As this method lacks native complex-domain FDR control, we apply the following real-value transformation with $\Re(\cdot)$ and $\Im(\cdot)$ denoting real- and imaginary parts respectively,
\begin{equation}
    \XB_{R} = \begin{pmatrix}
  \Re{(\XB)} & -\Im{(\XB)} \\
  \Im{(\XB)} & \quad \Re{(\XB)}
\end{pmatrix}, \quad \, \XB_{R} \in \mathbb{R}^{2n \times 2p}
\end{equation}
\begin{equation}
    \yB_{R} = [\Re(\yB) \quad \Im(\yB)]^\top, \quad \, \yB_{R} \in \mathbb{R}^{2n}
\end{equation}
and consequently $\betaB_{R} \in \mathbb{R}^{2p} $.
The transformation preserves the properties of complex-valued multiplication and conjugation.
After the determination of the set of active indices, the estimated support set of $\betaB$ is split in half, accounting for real and imaginary parts, respectively.
The final estimate of the support of $ \betaB $ is obtained using the set operations union, i.e., 
\begin{equation}
    \mathrm{support(\widehat{\betaB})} = \mathrm{support}(\Re(\widehat{\betaB})) \cup \mathrm{support}(\Im(\widehat{\betaB})) \, ,
    \label{eq:cknock_union}
\end{equation}
and intersection, i.e., 
\begin{equation}
    \mathrm{support(\widehat{\betaB})} = \mathrm{support}(\Re(\widehat{\betaB})) \cap \mathrm{support}(\Im(\widehat{\betaB})) \, .
    \label{eq:cknock_intersection}
\end{equation}
In the presented simulation, we set the cardinality $\| \betaB \|_{0} = 5$, $p = 1000$, $n = 300$.
Also, we set the target FDR $\alpha = 10\%$, and evaluate the variable selection performance for a linearly valued signal-to-noise ratio ($\SNR$) set $ \SNR \in \{0.1, 0.5, 1, 2, 5, 10 \}$.
The non-zero support of $\betaB \sim \exp{(\imagunit \cdot \mathcal{U}(0, 2\pi))}$ with $\imagunit = \sqrt{-1}$.
Additionally, $\XB \sim \CN(0, \mathbf{I})$ and the noise term $\varepsilonB \sim \CN(0, \sigma^{2} \mathbf{I})$.
For the \emph{model-X+} based \emph{cKnock} selector, the Knockoffs were generated as Gaussian Knockoffs, as any attempt to use second-order Knockoffs turned out to be computationally infeasible within the set processing bound of 24 hours.
Also, we used the approximate semi-definite programming solver, as any attempt to solve the problem as a true semi-definite programming problem was computationally infeasible.
As shown in Fig.~\ref{fig:complex_regression_fdr_plot}, the proposed \emph{CT-Rex} selector, indeed controls the target FDR $\alpha$ for each of the $\SNR$ levels.
The modified \emph{model-X+} \emph{cKnock} selector achieves this aim only when using the intersection (\ref{eq:cknock_intersection}). The \emph{CT-Rex} clearly outperforms the \emph{model-X+} based \emph{cKnock} methods in terms of the True Positive Rate (TPR).

% -------------------------------------------------------------------------------
\subsection{Single Snapshot Compressed Beamforming (CBF) based DOA Estimation}
\label{sec:single_snap_DOA}
\begin{figure}[ht]
    \centering
    \subfloat[FDR-control property (Target FDR $\alpha = 10\%$) - Homogeneous\label{fig:compressed_beam_form_fdr_plot}]{
        \includegraphics[width=0.47\linewidth]{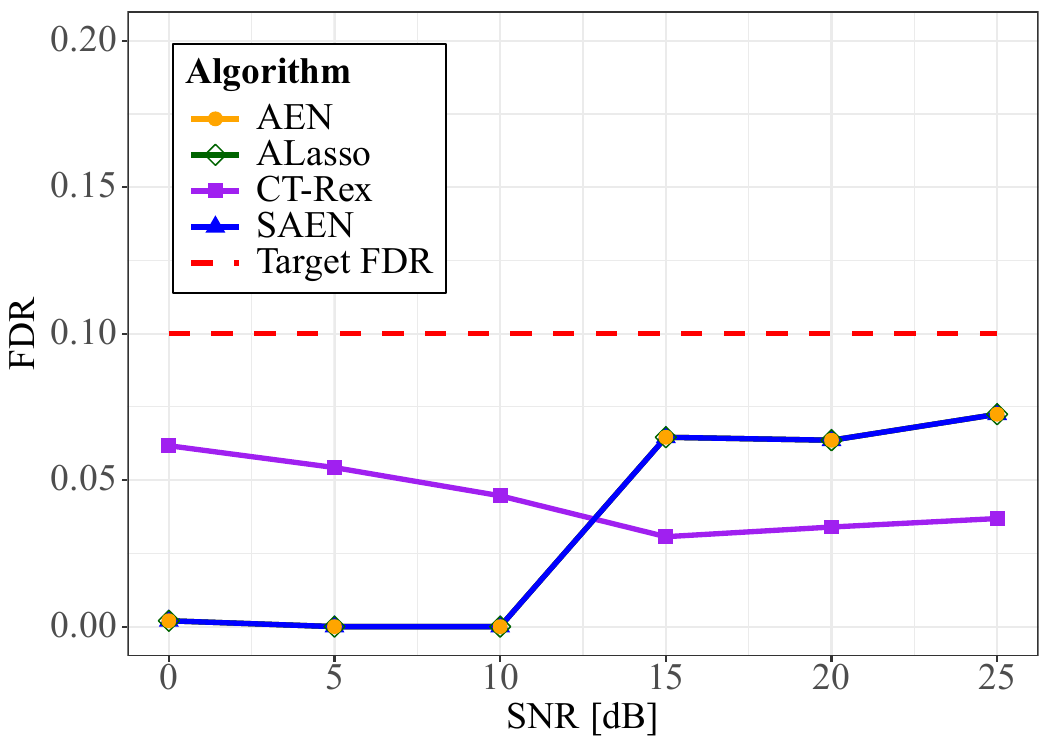}
    }
    \hfill
    \subfloat[FDR-control property (Target FDR $\alpha = 10\%$) - Heterogeneous sources.\label{fig:compressed_beamform_hetero_fdr_plot}]{
        \includegraphics[width=0.47\linewidth]{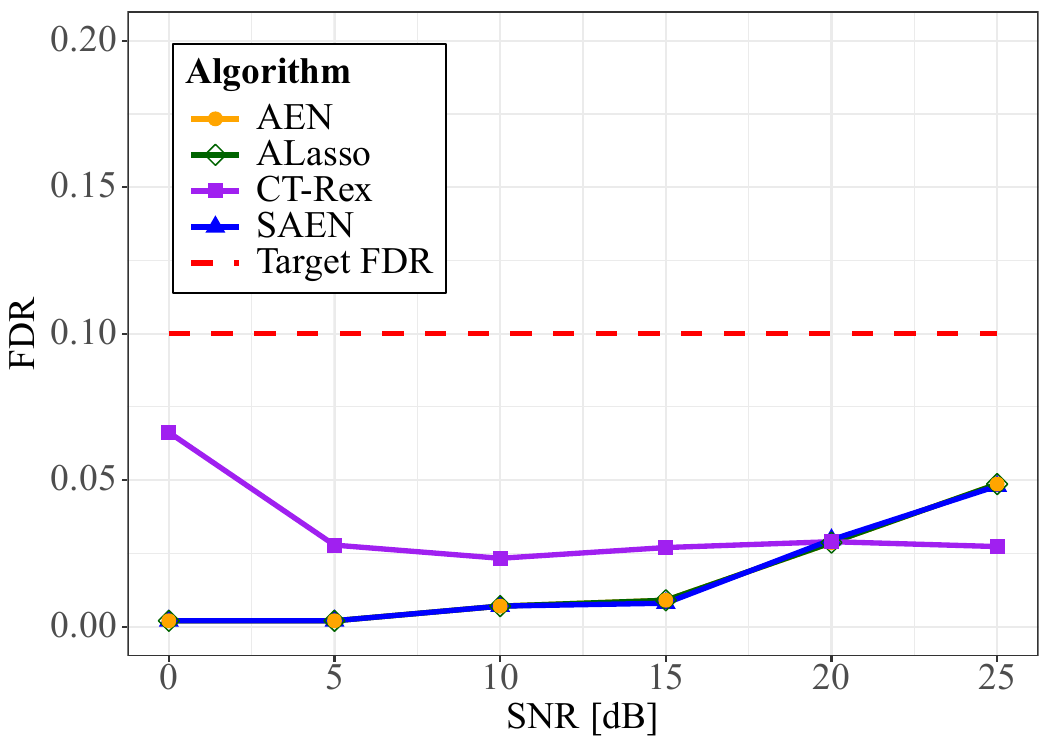}
    }
    \vspace{1em}
    \subfloat[TPR performance - Homogeneous sources.\label{fig:compressed_beam_form_tpr_plot}]{
        \includegraphics[width=0.47\linewidth]{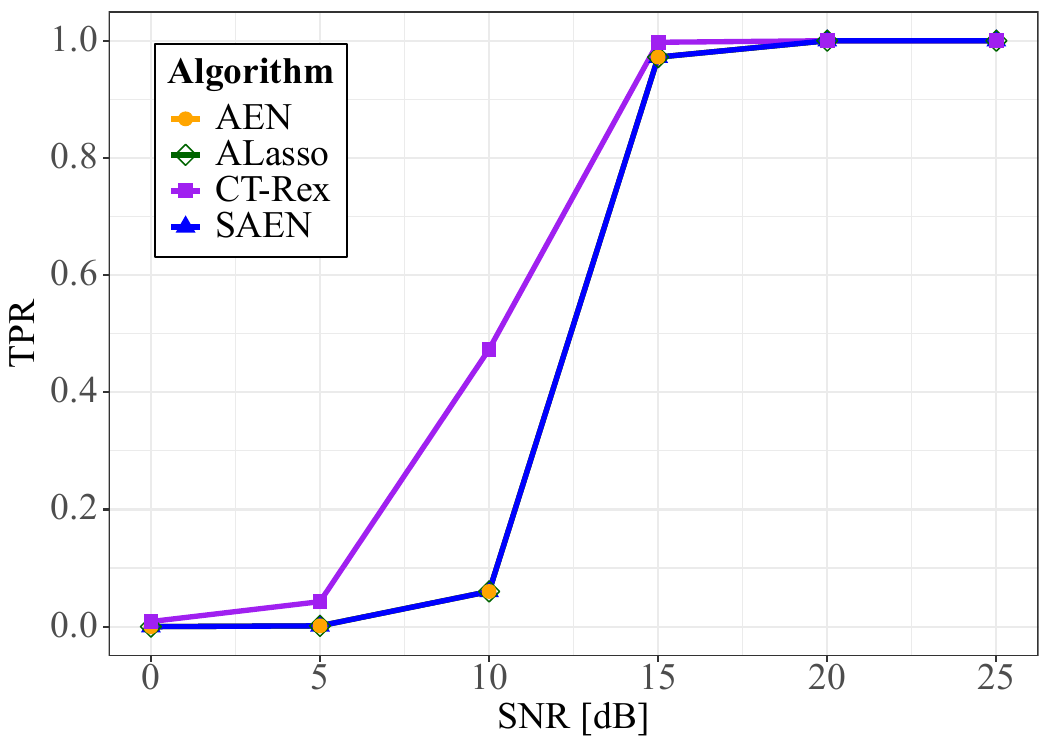}
    }
    \hfill
    \subfloat[TPR performance - Heterogeneous  sources.\label{fig:compressed_beamform_hetero_tpr_plot}]{
        \includegraphics[width=0.47\linewidth]{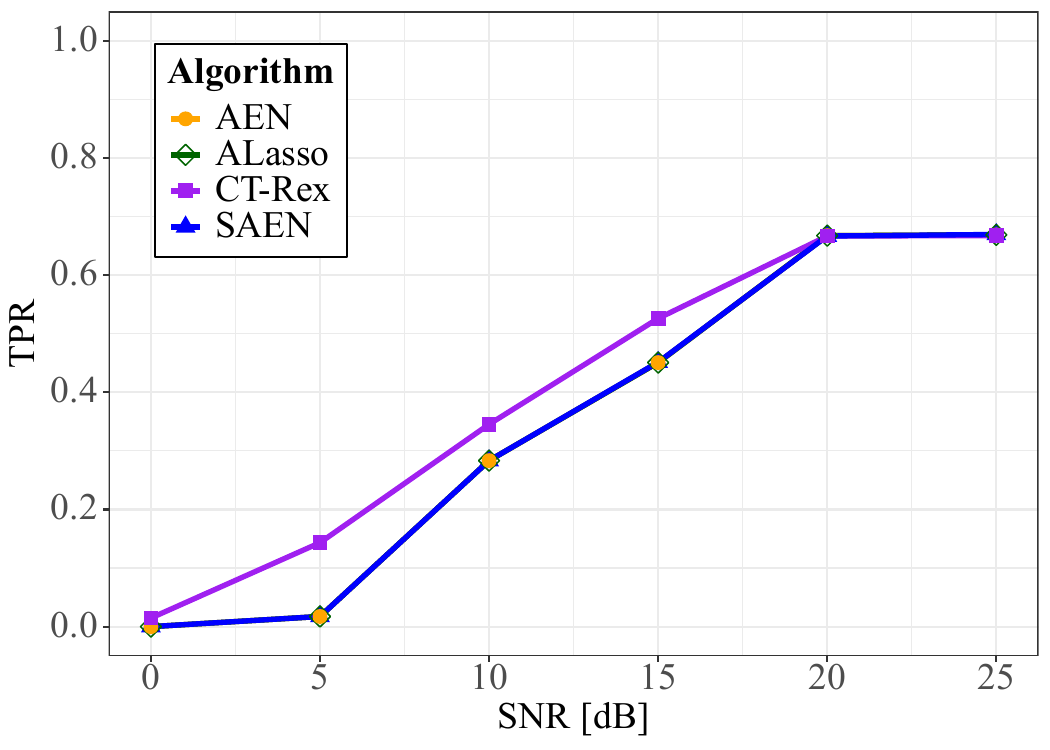}
    }
    \caption{CBF DOA estimation for varying SNR levels with homogeneous (left) and heterogeneous (right) source power.}
    \label{fig:compressed_beamforming_combined}
\end{figure}

The following scenario concerns the estimation of the direction of arrival (DOA) of source signals, a fundamentally important task, e.g., in radar and sonar systems.
The work of \cite{malioutov2005sparse} presented compressed sensing (CS) algorithms applied for DOA estimation, and demonstrated that the array output $\yB$ can be understood as a sparse linear model, utilizing a discretized DOA space.
Given the estimate of the support, the DOA estimates can be obtained by mapping them on the grid.
Note that the performance of compressed beamforming algorithms of \cite{malioutov2005sparse} depends on the grid size and spacing. A denser grid results in a large mutual coherence among the basis vectors and thus a poor recovery region for CS techniques follows.
Inspired by \cite{tabassum2016single, tabassum2018sequential, tabassum2018simultaneous}, we therefore follow a Lasso-type sparse regression approach and compare to their research results. To be comparable, we focus on the settings chosen in their papers. In particular, we assume narrowband processing, and we make the far-field assumption (i.e., propagation radius $\gg$ array size), resulting in a plane wave signal model.
% ------------------------------------
The sensor array is a uniform linear array (ULA) of $ M $ sensors aiming at DOA estimation for $\theta \in [-90^{\degreeSign}, 90^{\degreeSign})$ relative to the array axis.
The steering vector of a ULA with half-wavelength inter-element spacing for a source at angle $\theta$ is
\begin{equation}
    \mathbf{a}(\theta) = \frac{1}{\sqrt{M}} \left[1 \ e^{\imagunit \pi \sin(\theta)} \dots \ e^{\imagunit \pi (M - 1) \sin(\theta)} \right]^\top .
\end{equation}
For $Q < M$ sources from distinct DOAs $\{ \theta_{q} \}_{q=1}^{Q}$, a single snapshot ULA measurement can be modeled as
\begin{equation}
    \yB = \mathbf{A}(\theta) \stildeB + \varepsilonB \, ,
\end{equation}
where $\mathbf{A}(\theta) = [\mathbf{a}(\theta_1) \, \dots \, \mathbf{a}(\theta_Q)]$ is the steering matrix and $\stildeB \in \mathbb{C}^{Q}$ denotes the source support vector.
For DOAs on a uniform angular grid $\{ \theta_g \}_{g=1}^{G}$ over $[-90^{\degreeSign}, 90^{\degreeSign})$, the CBF model of \cite{tabassum2016single, tabassum2018sequential} applies with measurement matrix $\PhiB \in \mathbb{C}^{M \times G}$ as
\begin{equation}
    \yB = \PhiB \betaB + \varepsilonB \, ,
\end{equation}
with $\betaB \in \mathbb{C}^{G}$ being of cardinality $\| \betaB \|_{0} = Q$.
% ------------------------------------
We first present an example with $Q = 3$ homogeneous sources at fixed locations $(\theta_{1}, \theta_{2}, \theta_{3}) = (35^{\degreeSign}, 40^{\degreeSign}, 45^{\degreeSign})$, uniform source powers $(\eta_{\theta_{1}}, \eta_{\theta_{2}}, \eta_{\theta_{3}}) = (1.0, 1.0, 1.0)$, using $M = 80$ sensor elements and a grid resolution of $1^{\degreeSign}$.
We conducted $500$ Monte Carlo trials per algorithm and for each trial, the phase of the $q$th source signal is drawn randomly from a uniform distribution $\mathcal{U}(0, 2\pi) $.
The error term $\varepsilonB \sim \CN(0, \sigma^2 \mathbf{I})$, and we evaluated the $\SNR$ grid $\{0, 5, 10, 15, 20, 25 \}$ dB.
All methods, except the \emph{CT-Rex}, which does not require source enumeration, use the generalized information criterion according to \cite{tabassum2018simultaneous} to estimate the source number.
We report FDR and TPR in Figs.~\ref{fig:compressed_beam_form_fdr_plot}-\ref{fig:compressed_beamform_hetero_tpr_plot}, and the exact recovery performance in Table~\ref{tab:source_recoveries}.
As a second example, we present the situation of heterogeneous source powers $(\eta_{\theta_{1}}, \eta_{\theta_{2}}, \eta_{\theta_{3}}) = (0.3, 1.0, 0.04)$ for the source constellation as described before.
Fig.~\ref{fig:compressed_beamform_hetero_fdr_plot} and Fig.~\ref{fig:compressed_beamform_hetero_tpr_plot} demonstrate the TPR and FDR performances in this scenario.
Table~\ref{tab:source_recoveries} again reports the exact source recovery performance.
From the results, we can conclude that the FDR is controlled and the TPR of the \emph{CT-Rex} is slightly higher than that of its competitors in all SNR. 
However, this work is only a first step in FDR-controlled single snapshot DOA estimation.
Future research may focus on developing a dependency-aware version of the \emph{CT-Rex} in the spirit of \cite{machkour2024high} to address high correlations between steering vectors arising from closely spaced sources, where the competing methods perform well.
We also observed that for \emph{CT-Rex} the number of sensors must be high enough ($M \geq 80$ demonstrated good results).
Future work should investigate these points, considering off-grid solutions and also extend to multiple snapshot scenarios.

\begin{table}[htbp]
    \centering
    \caption{Exact source recoveries at varying SNR levels for homogeneous and heterogeneous source powers at a target FDR of $\alpha = 10\%$.}
    \label{tab:source_recoveries}
    \begin{tabular}{rrrrr}
    \toprule
    SNR Level & SAEN & AEN & ALasso & \emph{CT-Rex} \\
    \midrule
    \multicolumn{5}{c}{Homogeneous Source Powers} \\
    \midrule
    10 dB &   9 &   9 &   9 &  62 \\
    15 dB & 371 & 371 & 371 & 437 \\
    20 dB & 385 & 385 & 385 & 435 \\
    25 dB & 364 & 364 & 364 & 428 \\
    \midrule
    \multicolumn{5}{c}{Heterogeneous Source Powers} \\
    \midrule
    20 dB & 2 & 2 & 2 & 0 \\
    25 dB & 4 & 3 & 3 & 1 \\
    \bottomrule
    \end{tabular}
\end{table}

% ------------------------------------------------------------------------------------
\section{CONCLUSIONS}
\label{sec:Conclusions}
% Your conclusions content goes here.
% ------------------------------------
This work proposed the Complex-Valued Terminating-Random Experiments (\emph{CT-Rex}) selector, a fast FDR-controlling variable selector for complex-valued high-dimensional data.
The simulations confirmed the theory that the \emph{CT-Rex} selector controls a user-defined target FDR while maximizing the number of selected variables.
An important algorithmic contribution of this work is a complex-valued terminating LARS algorithm (\emph{CT-LARS}), which may be used as a forward selector, independent of FDR control.
We first benchmark the performance of the proposed method in complex sparse linear regression, comparing it to an adaptation of the \emph{Model-X+} Knockoff selector \cite{candes2018panning} for complex-valued data.
Then, we showcase a direction of arrival (DOA) estimation application.
Our results highlight the potential of FDR control in the complex domain and advocate for further research, including dependency-aware extensions.
The proposed \emph{CT-Rex} selector has wide applicability in fields using complex data, including DOA-based estimation, communications, mechanical engineering, medical imaging, and various areas of signal processing.

% ------------------------------------------------------------------------------------
\clearpage
\printbibliography
% ------------------------------------------------------------------------------------
\end{document}